\documentclass{jpsj-suppl}
\usepackage{txfonts} 
\usepackage{wrapfig}

\title{Power Spectrum Density of long-term MAXI data}

\author{Juri \textsc{Sugimoto}$^{1,}$$^{2}$, Tatehiro \textsc{Mihara}$^{1}$, Mutsumi \textsc{Sugizaki}$^{1}$, Motoko \textsc{Serino}$^{1}$,\\ Shunji \textsc{Kitamoto}$^{2}$, Ryousuke \textsc{Sato}$^{3}$, Yoshihiro \textsc{Ueda}$^{3}$, Shiro \textsc{Ueno}$^{4}$}

\inst{$^{1}$MAXI team, Institute of Physical and Chemical Research (RIKEN), Wako, Saitama 351-0198, Japan\\
$^{2}$Department of Physics, Rikkyo University, Nishi-Ikebukuro, Toshima, Tokyo 171-8501, Japan\\
$^{3}$Department of Astronomy, Kyoto University, Oiwake-cho, Sakyo-ku, Kyoto 606-8502, Japan\\
$^{4}$ISS Science Project Office, ISAS, JAXA, Sengen, Tsukuba, Ibaraki 305-8505, Japan}
	
\email{sugimoto@crab.riken.jp}

\recdate{July 15, 2013; accepted August 29, 2013; to appear in proceedings of The 12th Asia Pacific Physics Conference of AAPPS}

\abst{
Monitor of All-sky X-ray Image (MAXI) on the International Space Station has been observing the X-ray sky since 2009 August 15. 
It has accumulated the X-ray data for about four years, so far.
X-ray objects are usually variable and their variability can be studied  by the power spectrum density (PSD) of the X-ray light curves.
We applied our method to calculate PSDs of several kinds of objects observed with MAXI.
We obtained significant PSDs from 16 Seyfert galaxies.
For blackhole binary Cygnus X-1 there was a difference in the shape of PSD between the hard state and the soft state.
For high mass X-ray binaries, Cen X-3, SMC X-1, and LMC X-4, there were several peaks in the PSD corresponding to the orbital period and the superorbital period.
}

\kword{MAXI, PSD, HMXB, AGN, Cygnus X-1}

\begin{document}
\maketitle

\section{Introduction}


X-ray stars are variable in its intensity with many time scales. 
Sometimes we can obtain specific periods and sometimes we can find particular variability characteristic, such as red noise. 
They should relate to their emission properties, such as sizes of their emission region, a transition of accretion discs, 
precession of accretion disks and so on, as well as binary orbital periods.  
We study the characteristics of time variations of several kinds of objects obtained with the Monitor of All-sky X-ray Image (MAXI), with the power spectrum analysis.

\section{Observation}
MAXI onboard the International Space Station (ISS) was launched in 2009 July\cite{matsuoka}.
The ISS with MAXI orbits the earth in 92 minutes, and MAXI scans the objects in the all sky once in the period. 
The MAXI has already reported more than one hundred of transients ( http://maxi.riken.jp/top/doc/maxi\_atel.html ).
The observed results are immediately released through the Internet, promoting immediate follow-up observations with telescopes around the world.
The energy band of MAXI/GSC is 2$-$30 keV.
We use the MAXI observed data in four energy bands : 2$-$20 keV, 2$-$4 keV, 4$-$10 keV and 10$-$20 keV 
taken from the MAXI homepage ( http://maxi.riken.jp/ ).\\

\section{PSD analysis}
The X-ray light curve is converted to the PSD by discrete fourier transformation (eq.\ \ref{eq:FTcos})\cite{scagle}.
  \begin{equation} 
F_{\rm c}(f) = \frac{1}{N-N_{gap}}\sum_{i=1}^{N}y_{i}\cos(2\pi ft_{i})\label{eq:FTcos}, \ \ \ \ \ \ 
F_{\rm s}(f) = \frac{1}{N-N_{gap}}\sum_{i=1}^{N}y_{i}\sin(2\pi ft_{i})
  \end{equation}
where $y_{i}$ is the intensity (photons s$^{-1}$ cm$^{-2}$), $N$ is number of data after filling the gap, $N_{gap}$ is number of data in the gap, 
and $t_{i}$ is time in MJD.
Since the sampling interval is equal, $t_{i}$ is par day or 90 minutes.
$f$ is the frequency which is a discrete value for multiples of $\Delta f$=$\frac{1}{T}$ 
where $T$ is total time of the observation.
Because the filled data has no power from the source, we divided by (N-N$_{gap}$) instead of N.
The PSD is the sum of squares of $F_{\rm c}$ and $F_{\rm s}$ as
\begin{equation}
  PSD(f) = \frac{T}{2}\{F_{\rm c}^2(f) + F_{\rm s}^2(f)\}
\label{eq:PSD}
\end{equation}

There are some gaps in the observation because of sun angle limitation or the ISS's rotational axis.
However, the data set needs to be continuous for fourier transform.
We evaluated the correctness of the methods by simulations.
We simulated by three method.
First, we interpolated 0 in the gap.
In this method PSD has too much power.
Next, we interpolated by with the mean value of the light curve.
This method makes abrupt steps small enough, 
and the fake power of the PSD decreased.
In the end, we found the following method is the best. 
We calculated the averages of several data points at both ends of the gap,
and interpolated linearly in the gap.

\begin{figure*}[htb]
 \begin{center}
   \includegraphics[scale=0.25]{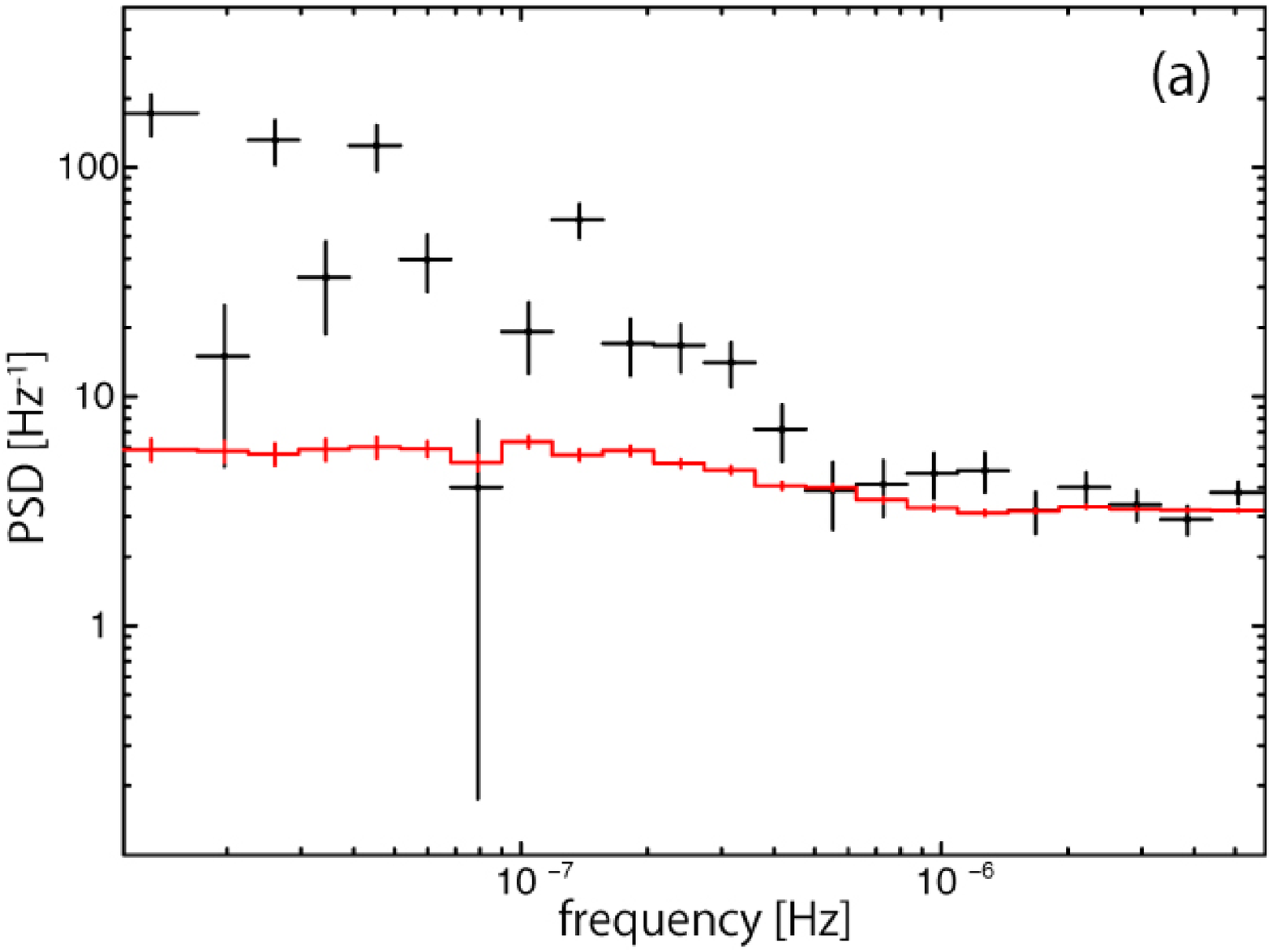}
	\hspace{30pt}
   \includegraphics[scale=0.25]{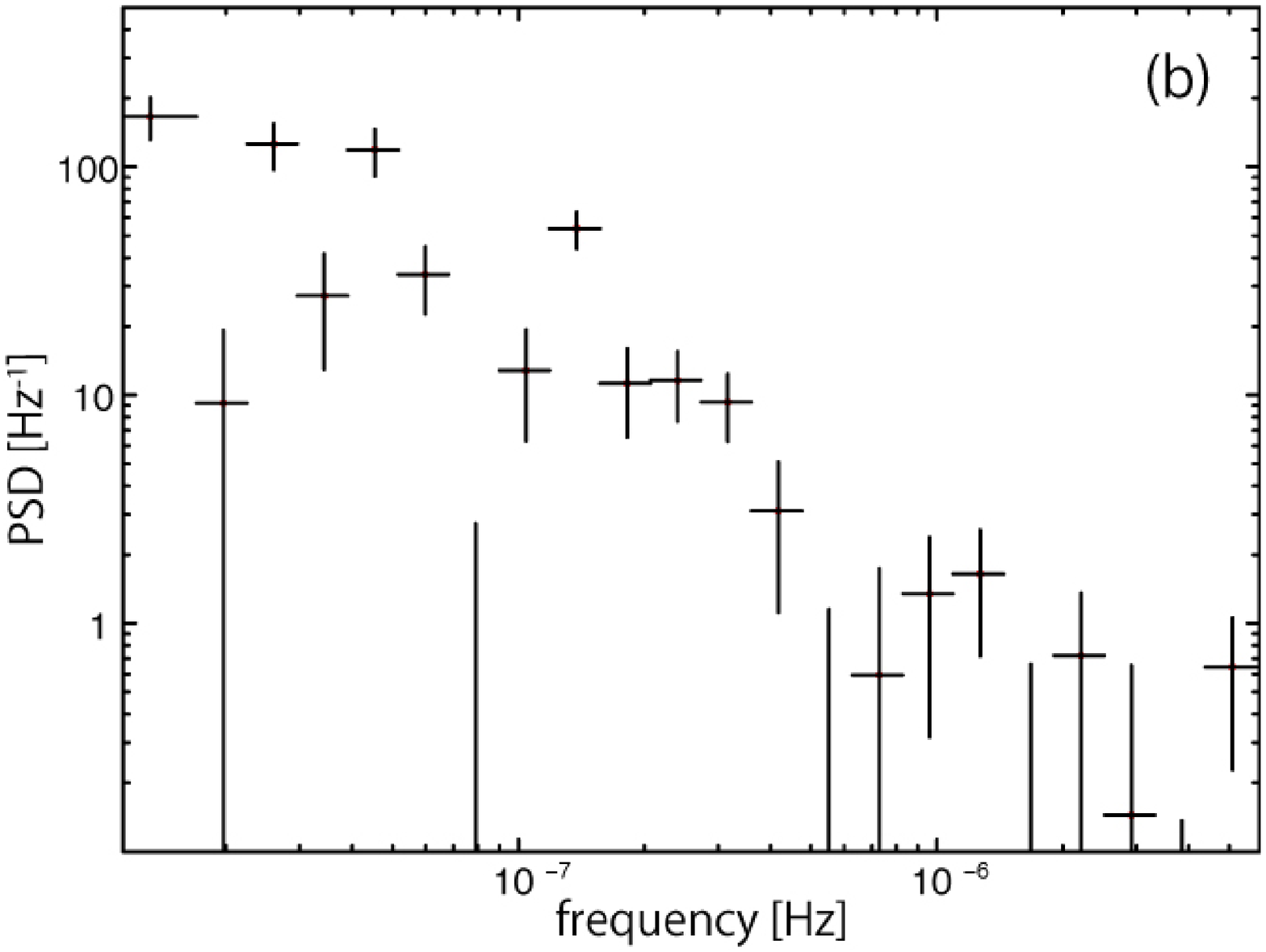}
 \end{center}
 \caption{
(a) The PSD of Centaurus A obtained by calculating from the observed data.
The red line shows the simulated Poisson level.
(b) The true PSD of the source after subtracted the Poisson level.
}
\label{fig:psd+pl}
\end{figure*}

The effect of linear interpolation is evaluated by simulation.
The observed data also have the statistical error.
Since it is random, it appears in PSD as a constant value.
It is called the Poisson level.
To obtain a true fourier spectrum of the source, we should subtract the Poisson level from the observed PSD. 
However, the MAXI data includes various quality data due to the observation conditions, for example the effective area is different in each sampling.
We estimated the Poisson level by numerical simulation in the following way.
First, we calculated the average intensity of an object and generated a random number with a poisson distribution for both the source integrated area and the background integrated area.
We made the simulated light curve with statistical fluctuation, taking into account of the elevation angle of the source in the field of view.
Then we made a hundred of light curves and PSDs, and determined the Poisson level with the effect of linear interpolation by averaging them.
Fig \ref{fig:psd+pl} is the PSD and simulated Poisson level of the Seyfert galaxy Centaurus A.
In Fig \ref{fig:psd+pl} (a) the black points show the observed data, and the red line shows the simulated PSD.
The effect of linear interpolation appears as a small rise above the Poisson level in low frequency.
The resultant PSD subtracted the simulated PSD is shown in Fig \ref{fig:psd+pl} (b).

\begin{table}[htb]
\begin{center}
\caption{
Periods of HMXBs obtained from PSD in 4-10 keV with MAXI/GSC.
We wrote bin width = $\Delta f$ as an error.
}
\label{tb:HMXB}
\begin{tabular}{llll}
\hline
Source & span [MJD] & $P_{{\rm orb}}$ [day] & $P_{{\rm sup}}$ [day]\\
\hline
Cen X-3 & 55058-56363 & 2.09 $\pm$ 0.01 & 93.3$\sim$435.1 \\
SMC X-1 & 55058-56400 & 3.90 $\pm$ 0.01 & 52.1$\sim$81.5 \\
LMC X-4 & 55058-56450 & 1.40 $\pm$ 0.01 & 30.5 $\pm$ 0.8 \\
\hline
\end{tabular}
\end{center}
\end{table}
\begin{figure*}[htbp] 
\begin{center}
 \hspace*{-8mm}
   \includegraphics[scale=0.27]{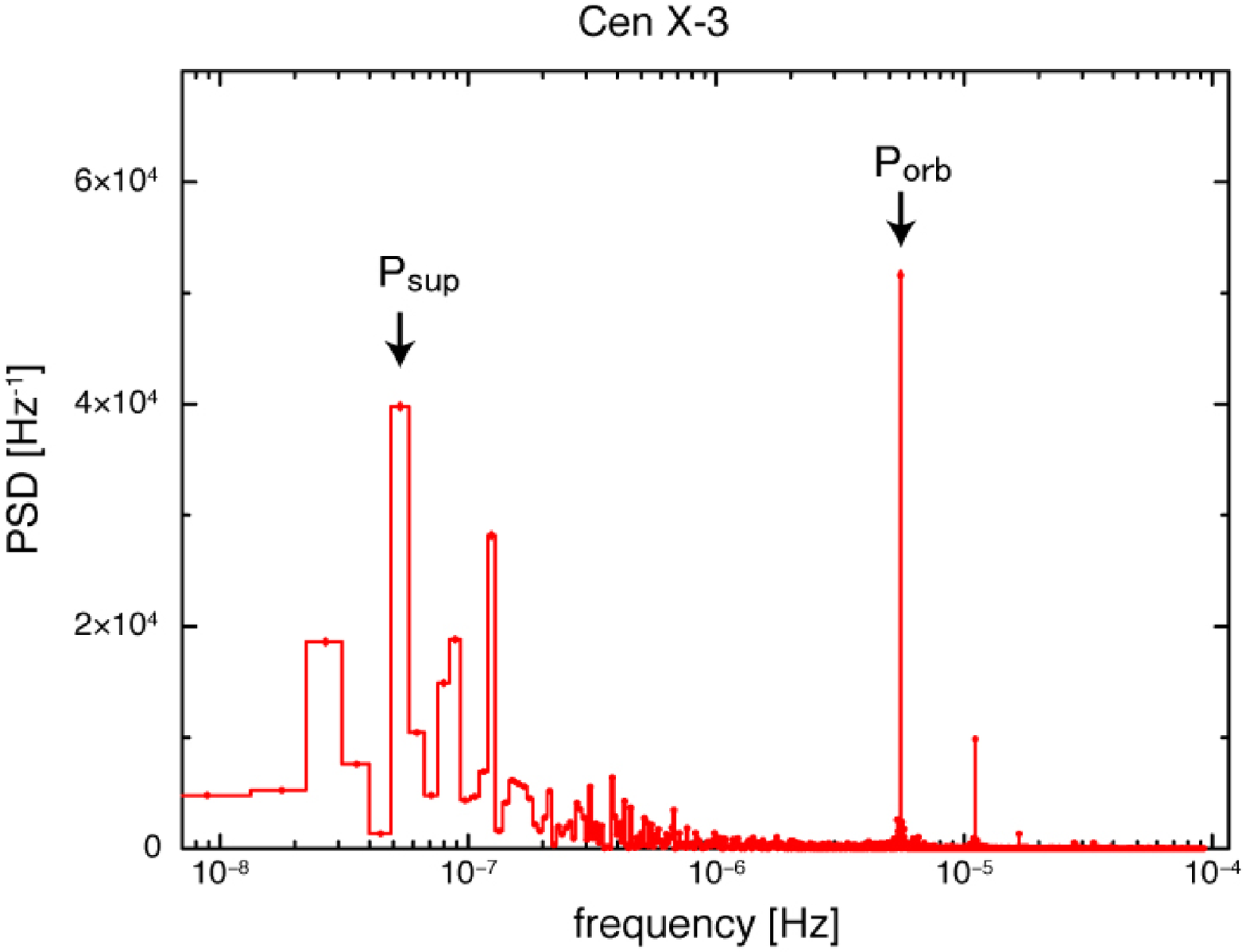}
   \includegraphics[scale=0.27]{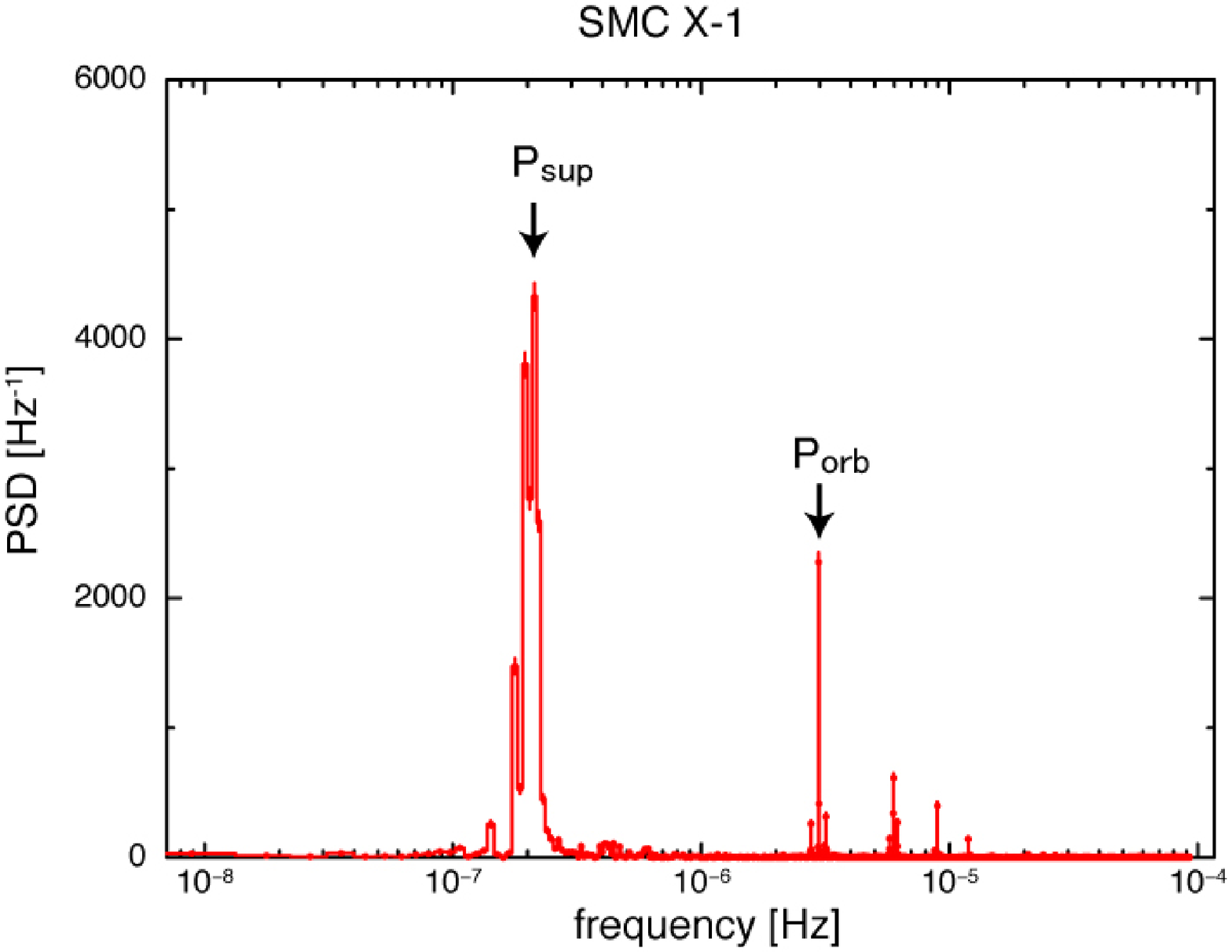}
   \includegraphics[scale=0.27]{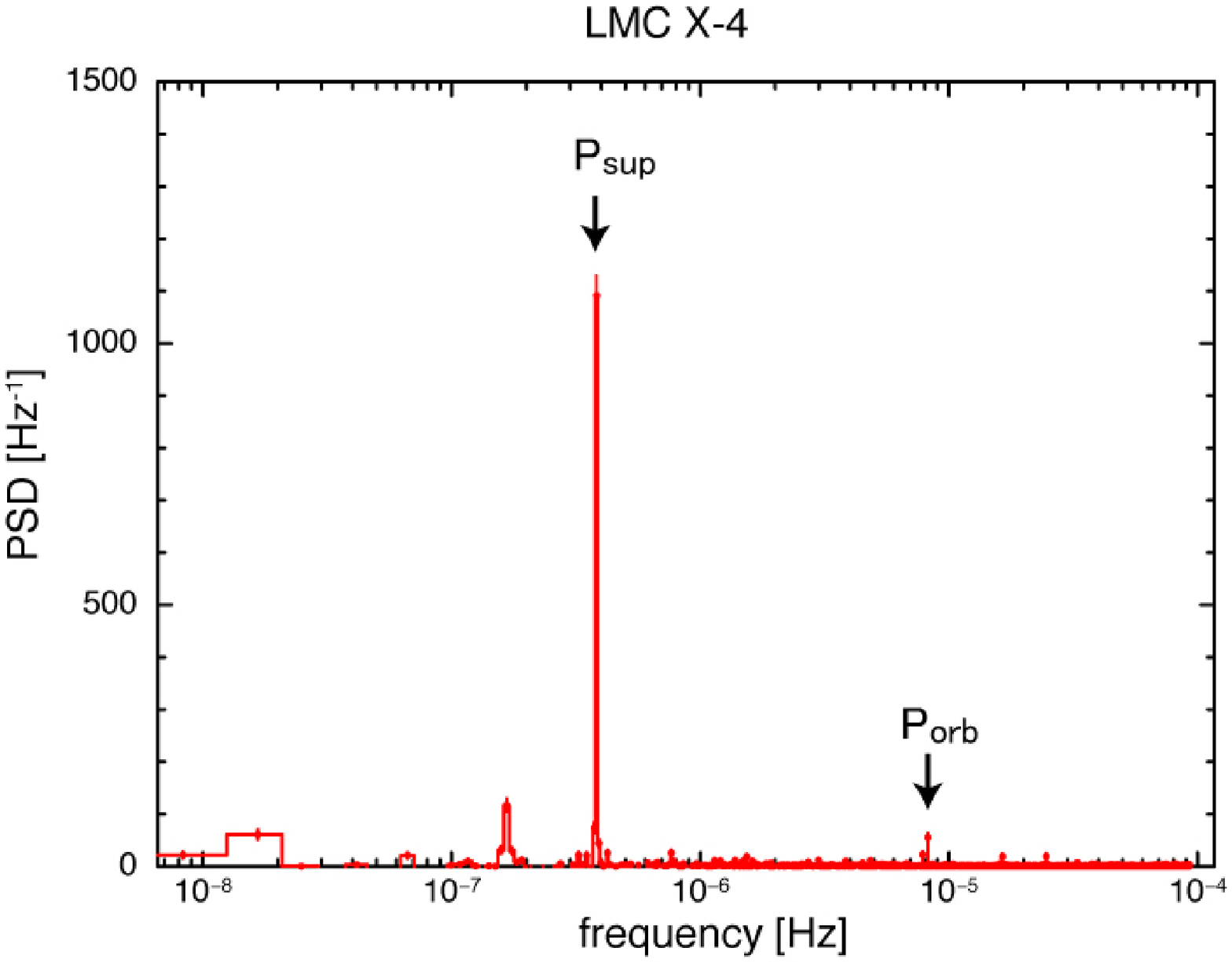}
 \end{center}
 \caption{
The PSDs of Cen X-3, SMC X-1 and LMC X-4.
The values are listed in Table \ref{tb:HMXB}.
}
\label{fig:HMXB}
\end{figure*}

\section{Results}
\subsection{Periods of HMXBs}
We calculated the PSD from the MAXI data in 4-10 keV per 90 minutes of three HMXBs whose periods are already known.
We detected the orbital periods and the superorbital periods from Cen X-3, SMC X-1, and LMC X-4.
The results are shown in Table \ref{tb:HMXB} and Fig \ref{fig:HMXB}, which are consistent with the previous research \cite{kotze}.
LMC X-4 has a stable superorbital period of 30.5$\pm$0.8 days.
Whereas, SMC X-1 and Cen X-3 have several peaks in tens of days and in hundreds of days.
This is because that the superorbital periods are unstable in some objects as previous research \cite{kotze} reported.

\subsection{NPSD of Cygnus X-1}

\begin{wrapfigure}[17]{r}{8.3cm}
 \begin{center}
  \vspace{-3.5\baselineskip}
   \includegraphics[scale=0.4]{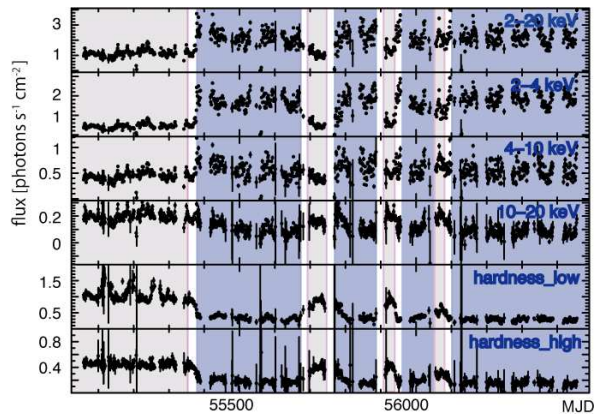}
\label{fig:hr}
 \caption{
The light curve and the hardness ratio of Cygnus X-1.
The area with light gray is the hard state, and the area with dark gray is the soft state.
}
\end{center}
\end{wrapfigure}
Cygnus X-1 is a well-known blackhole binary.
It shows a two distinct states in energy spectrum : the hard state and the soft state.
The hard state is a less bright state and has a power law dominated spectrum.
The standard accretion disk is truncated at some radius.
The disk expands to become optically thin at inner part, which produces a power law spectrum.
The soft state is a bright state dominated by a disk blackbody.
The standard disk plunged to the innermost stable circular orbit.
Cygnus X-1 stayed in the hard state in most of the time.
However, during the MAXI observation Cygnus X-1 turned into a soft state (Fig 3).
We focused on the PSDs of the both states of Cygnus X-1.

\begin{figure*}[thbp]
 \begin{center}
\vspace{-2\baselineskip}
   \includegraphics[scale=0.65]{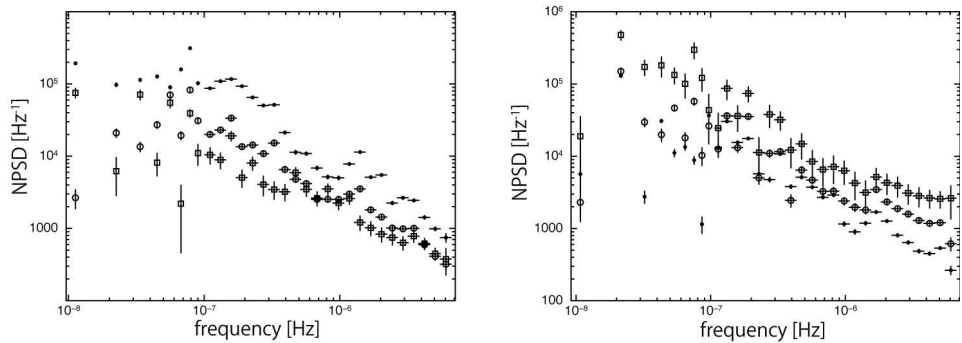}
 \end{center}
\vspace{3mm}
 \caption{
The results of the NPSD in three energy bands.
The left figure is in the hard state, and the right one is in the soft state.
The filled circles are in 2-4 keV, the open circles are in 4-10 keV and the open squares are in 10-20 keV.
}
\vspace{-1\baselineskip}
\label{fig:cyg}
\end{figure*}

First, we divided the whole period into the hard and the soft states by calculating the hardness ratio.
Cygnus X-1 is in the hard state for about ten months from the beginning. 
After that, it changed to the soft state and continued for another ten-month.
Afterwards it repeated the state change in every 1$\sim$3 months. 
After 56107 MJD it is in the soft state.
We examined the variation by making PSD in each state.
For calculating the PSD in the hard state, we used only the data in the hard state filling the gap of the soft state by the average intensity of the hard state.
We discarded the data at the transition of these states.
Since the PSD value is large when the source is strong, we normalized the PSD by dividing by the square of the average intensity (NPSD).
The result is shown in Fig \ref{fig:cyg} in 2-4 keV, 4-10 keV and 10-20 keV bands.

In the hard state, the power of the NPSD in 2-4 keV is larger than those in 4-10 keV and 10-20 keV.
In the soft state, the power of the NPSD in 10-20 keV is larger than those in other energy bands.
Reig et al. (2002) analyzed the PSD of Cygnus X-1 using the {\it RXTE}/ASM data in 1.3-12.2 keV.
Cygnus X-1 was almost in the hard state in their observed span (50139-52200 MJD).
Our results in 2-4 keV and 4-10 keV in the hard state are consistent with Reig et al. (2002).
We obtain the NPSD of the hard state in the 10$-$20 keV band as well as those of the soft state in all the bands for the first time.
Our results show that the disk-blackbody component is variable in the hard state suggesting that the inner radius of the standard accretion disk is changing more,
and that the power-law component is more variable in the soft state suggesting that the Comptonizing corona is changing more.

%
%
%

\section{Conclusion}
We calculated the broadband PSD with long-term MAXI data in order to fill the observational gaps.
We calculated the averages at several data points of both ends of the gap and interpolated the gap linearly. 
We determined the Poisson level with the Monte Carlo simulation to obtain the true PSD of the source.

From the PSD analysis, we obtained the orbital period and the superorbital period of three HMXBs, Cen X-3, SMC X-1 and LMC X-4.
We confirmed that the superorbital period of LMC X-4 is stable, while those of Cen X-3 and SMC X-1 are unstable.
We investigated the NPSD of Cygnus X-1 in the hard state and in the soft state.
In the hard state, the soft band is the most variable whereas in the soft state the hard band is the most variable. 
It would represent variable part of the system, which are inner radius of the accretion disk and the Comptonizing corona, respectively.


\end{document}